# Integration of Large Vision Language Models for Efficient Post-disaster Damage Assessment and Reporting

Zhaohui Chen [a], Elyas Asadi Shamsabadi [a], Sheng Jiang [b,c], Luming Shen [a],

Daniel Dias-da-Costa [a,*]

[a] *School of Civil Engineering, Faculty of Engineering, The University of Sydney, Sydney, NSW 2006, Australia*

[b] *The National Key Laboratory of Water Disaster Prevention, Hohai University, Nanjing 210098, China*

[c] *College of Water Conservancy and Hydropower Engineering, Hohai University, Nanjing, 210098, China*

*\*Corresponding author*

**Abstract:**

Traditional natural disaster response involves significant coordinated teamwork where speed and efficiency are key. Nonetheless, human limitations can delay critical actions and inadvertently increase human and economic losses. Agentic Large Vision Language Models (LVLMs) offer a new avenue to address this challenge, with the potential for substantial socio-economic impact, particularly by improving resilience and resource access in underdeveloped regions. We introduce DisasTeller, the first multi-LVLM-powered framework designed to automate tasks in post-disaster management, including on-site assessment, emergency alerts, resource allocation, and recovery planning. By coordinating four specialised LVLM agents with GPT-4 as the core model, DisasTeller autonomously implements disaster response activities, reducing human execution time and optimising resource distribution. Our evaluations through both LVLMs and humans demonstrate DisasTeller's effectiveness in streamlining disaster response. This framework not only supports expert teams but also simplifies access to disaster management processes for non-experts, bridging the gap between traditional response methods and LVLM-driven efficiency.

# 1. Introduction

**Natural disasters** consist of various catastrophic events, including earthquakes, hurricanes, floods, and wildfires. These events can cause extensive damages, loss of life, and have long-term impacts on societies and economies. For instance, earthquakes can cause buildings to collapse, trapping people under rubble, while hurricanes and floods can result in drowning and physical damage. These natural disasters also lead to extended social challenges, such as the disruption of community cohesion and the psychological effects of trauma and loss. According to the report from United Nations office, the overall global economic losses from natural disasters in 2023 were estimated to be US 250 billion dollars, which roughly equals the entire gross domestic product of New Zealand [1]. Beyond immediate losses, disasters would have long-lasting economic effects, including reduced economic growth, increased poverty, and disruption to markets and industries. The frequency and intensity of natural disasters have been rising due to factors such as climate change, urbanisation, and environmental degradation, making the need for effective post-disaster management even more critical.

**Post-disaster management** involves activities including disaster response and post-disaster recovery aimed at mitigating the impacts of natural disasters. The Sendai Framework of the United Nations for Disaster Risk Reduction (2015–2030) highlights the crucial role of enhanced disaster risk governance in ensuring an effective response and promotes the concept of 'Building Back Better' in sustainable recovery, reconstruction, and rehabilitation efforts [2]. Traditionally, the entire process of disaster response and post-disaster recovery requires broad collaboration and teamwork [3,4]. As shown in Fig. 1, the first step in the disaster response process involves the post-disaster on-site assessment through an expert team, which would focus on evaluating the structural integrity of buildings, transportation networks and other critical infrastructures. Based on the assessment of the expert team, the alerts team is activated. The role of the alerts team is to issue emergency alerts to the public, the emergency service team, and government agencies. These alerts are crucial for ensuring that all parties are aware of the situation and can act accordingly. Simultaneously, the emergency services team is mobilised to provide immediate relief to affected populations and determines the number and locations of relevant facilities based on the information provided by the alerts team and the expert evaluations. The assignment team works closely with other teams and is tasked with distributing human resources for emergency service construction, determining where and how resources should be assigned depending on the other three teams' reports. In addition, an essential component of the disaster response is keeping the affected communities informed. This includes advising them about the ongoing operations, where they can access shelter and medical care, and any other relevant updates. Effective communication helps to manage public expectations, reduce panic, and foster cooperation among affected communities. After the response phase, the government and policymakers could rely on updates from the assignment team to make decisions about rebuilding strategies and long-term resilience measures.

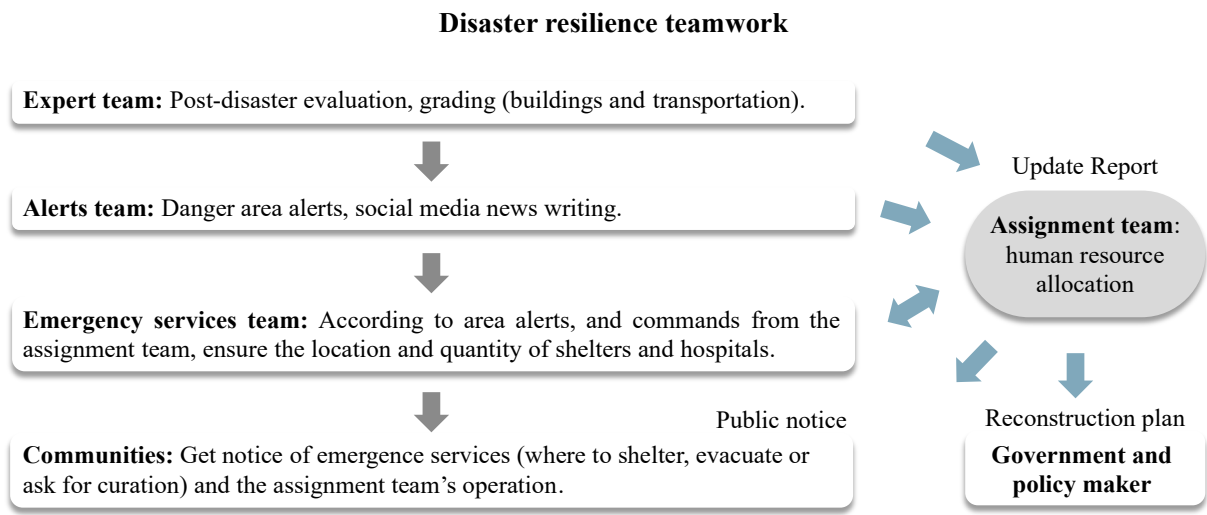

Fig. 1 The schematic of disaster response human teamwork

Generally, the disaster response work would last several weeks due to the limitation of human execution and mobility [5,6], and the response process would be even slower in underdeveloped areas due to insufficient support and infrastructure investment [7], inevitably leading to vast losses. In all periods, enhancing the speed of disaster response efforts and ensuring marginalised groups are not left behind are widely recognised as crucial for reducing disaster-related losses and saving human lives [8,9]. Current disaster management research focuses on disaster-related factors analysis under specific scenarios [10-12]. Math equations or simple machine learning models cannot handle the complex scenarios in post-disaster management, let alone simulate human collaborations to reduce labour work, accelerate the response, or even mitigate the disaster response disparities between developed and underdeveloped areas.

As **artificial intelligence (AI)** technology develops, it will provide the possibility to better emulate human intelligence, thus rendering it possible to assist or even replace humans to improve efficiency. One of the most significant breakthroughs in AI has been the development of large vision language models (LVLMs**).** These models, which include OpenAI's GPT-4 series [13], are trained on vast amounts of image and text data, enabling them to generate human-like text, understand image and context, and even perform complex tasks like summarisation and conversation. Large vision language models are built on neural networks that mimic the structure of the human brain allowing them to process and generate language in a manner that is increasingly difficult to distinguish from human output. Their ability to understand and generate language has made them valuable tools in various areas that require the processing and interpretation of large volumes of text or images [14]. As LVLMs are further developed, they are becoming more integral to the way humans interact with technology, driving innovations that will shape the future of communication and information processing.

Given the development of powerful LVLMs, utilising multi-LVLMs instead of human teams to execute the disaster response teamwork autonomously could accelerate the complete procedure, minimise disaster-related losses and narrow the disaster response gap between developed and underdeveloped areas. Unlike humans, LVLMs can operate continuously without fatigue across regions with different levels of development, carry out team tasks and interactions almost instantly, integrate data from historical data quickly and provide consistent analysis, avoiding the different levels of experience and subjective human judgement [15]. To explore the possibility of this concept, we propose here a multi-LVLM-powered disaster response framework, DisasTeller, designed to streamline the collaboration of different teams in post-disaster management. DisasTeller operates by prompting four independent LVLM agents (GPT-4o in our experiment) with specific task instructions and implementing three assistant tools to complete the process. The LVLM agents are provided with the descriptions of their roles, the utility of defined tools, and the details of expected input and output. The 'expert team' agents first analyse on-site local post-disaster and global region images, then share their information with the other three teams. The 'alerts team' and 'emergency team' agents reason the output of the 'expert team' and then generate the alerting news and emergency services report, respectively. After evaluating the outputs of the previous three teams, the 'assignment team' agents finalise the human resource allocation command, public announcement and the future reconstruction plan. DisasTeller is designed as an engine that can be further developed and easily adapt to new applications by introducing new agents and tools. DisasTeller could act as an aid for post-disaster response teams and simplify access to accurate post-disaster management knowledge for non-experts by providing an easy-to-use interface. We assess the reasonability of all six output reports from the agents of the four teams, including the concise summary of the disaster from the 'expert team', the alerting news from the 'alerts team', the emergency services report from the 'emergency team', and another three key reports from the 'assignment team'.

## 2. Results and discussions

### 2.1 Autonomous disaster response

To validate the autonomous process of DisasTeller, a case study of earthquake disaster was conducted in Fig. 2, utilising the data from previous literature [16]. We define LVLM agents by prompting task descriptions for several independent GPT-4o models with assigned tools, as shown in Fig. 2a, before the execution of DisasTeller. Like the initial action of the experts' on-site inspections for the post-disaster response, the input of the framework starts with local on-site post-disaster and global map images. As shown in Fig. 2b, we manually defined the affected area by earthquake as the Wajima City in Japan, and the locations of the local-view images as 'Wajima Drama Memorial Hall', 'Hama Street', 'Concrete Bridge', 'Central Nishikigawa Street', 'North Asaichi Street' and 'South Central Asaichi Street'. DisasTeller first processes the local-view images for the location and visible damage descriptions. Utilising 'file search tool', DisasTeller reads the local technical guideline file 'European Macroseismic Scale (EMS98)' [17] for the damage evaluation and grading (G1 – G5, from slight damage to very heavy damage). The output of the post-disaster damage grading in Fig. 2b matches the local-view images well, demonstrating the successful collaborations of LVLM agents to analyse different disaster information. After the damage evaluation of local areas, DisasTeller deploys the 'map annotation tool' to search and annotate the damage grades in the relevant locations of the global map image, thus generating the alert map to visualise the overall situation in this disaster area. This alert map is then further reviewed by other LVLM agents in DisasTeller to generate different reports (partial output of reports shown in Fig. 2c). The first report is generated by the 'expert team' in DisasTeller to briefly summarise the local-view disaster and their damage grades. After inspecting the 'expert team' report and the alert map, the 'alerts team' in DisasTeller starts reasoning and generating alert news, providing information and safety alerts of existing dangerous areas (e.g., 'Fire-Damaged Areas: Particularly around Asaichi Street and near the Wajima Morning Market, these zones have unstable structures and debris. Avoid these areas due to potential fire and structural hazards'). Analysing the reports of previous agent teams and taking reference to the historical disaster data from the internet through 'web search tool', 'emergency team' ensures the areas requiring immediate recovery and emergency services (e.g., 'Concrete Bridge: Immediate assessment and restoration are crucial to reestablish vital transport links'). Afterwards, the 'assignment team' integrates the information from previous three reports and disaster archives obtained online via 'web search tool' to generate the human resource allocation report (e.g., 'Wajima Drama Memorial Hall: This area will host a large emergency shelter. It requires a team of 20 medical personnel, including 5 doctors and 15 nurses, supported by 10 logistics personnel'), public notice report (e.g., 'We are coordinating with governmental agencies and non-governmental organizations to ensure a comprehensive response'), and future reconstruction plan (e.g., 'Based on previous case studies, an estimated budget of approximately $1 billion is needed to cover structural repairs'). DisasTeller saves the intermediate outputs of LVLM agent teams in memory and allows LVLM agents to reuse these intermediate outputs and defined tools provided they have access, thus automating the whole disaster response process by leveraging the reasoning capabilities of LVLMs.

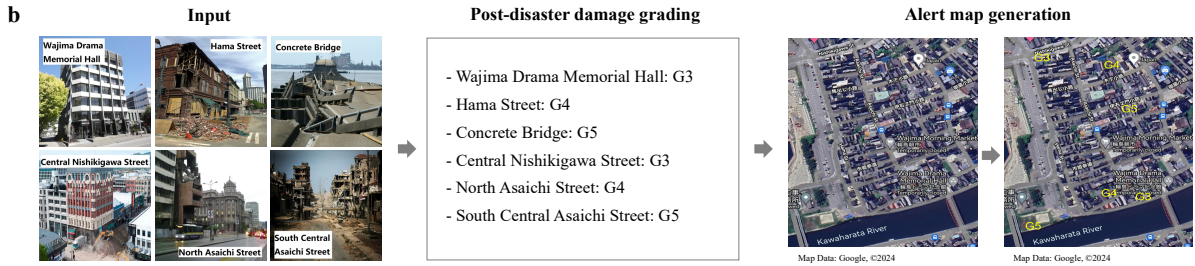

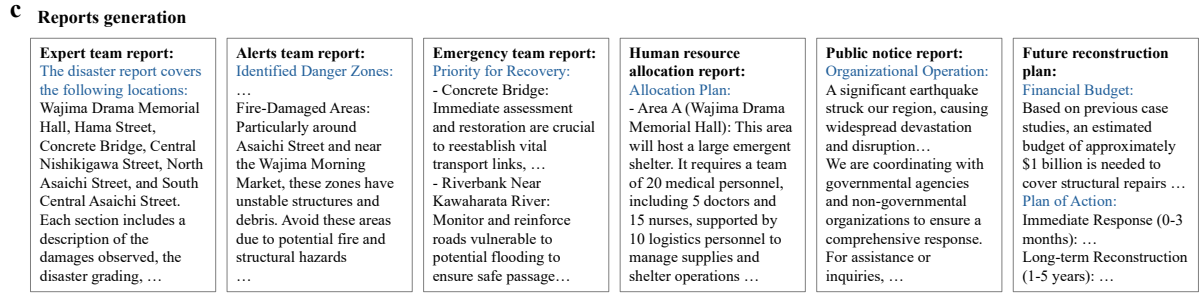

Fig. 2 The experimental process of our framework: a. the descriptions for tasks of agents and tools assigned; b. the running process of DisasTeller and intermediate tasks; c. the final output reports of agents in DisasTeller.

Reliable outputs of LVLM agents are critical for effectively executing DisasTeller. Yet, the quality of LVLMs' outputs highly relies on the prompt inputs for the LVLM agents, and even the same instruction prompt can lead to different outputs [18,19]. Designing prompts to achieve optimal results is an iterative process. Since maintaining identical and best outputs of LVLM agents is complex, LVLM agents in DisasTeller are restricted to output contents with designated format templates when defining task prompts, thus providing relatively controlled results. This case study demonstrates DisasTeller's ability to autonomously carry out post-disaster response tasks through LVLM agents' successful simulation of the coordination of disaster management teams.

## 2.2 Agent output results evaluation

In recent years, integrating LVLMs into machine learning frameworks has accelerated the discoveries in different scientific areas [14,20]. Yet, LVLMs can generate incorrect or deceptive contents and there are short of standardised evaluation metrics, posing a great challenge to assess the output quality of LVLM-based frameworks. Considering the previous studies [14,20], we conducted five independent runs of DisasTeller and implemented the comprehensive assessment with a combination of the LVLM (GPT-4o) and human inspection. The intermediate process tasks of DisasTeller are evaluated manually due to their simplicity, and the six output reports of DisasTeller are assessed by both LVLM (GPT-4o) and human inspection, considering their complexity. GPT-4o is prompted to act as a post disaster management expert (which is called EvaluatorGPT [21]) and has access to the post-disaster on-site images for more accurate assessment. The prompt for EvaluatorGPT is modified to grade the output reports of DisasTeller based on the coherence and consistency of sentences and the accuracy of claims. Each evaluation of EvaluatorGPT requires a detailed explanation of the report's weakness to support the corresponding grade. In addition, we manually inspect DisasTeller's outputs and utilise the same rules to provide grades as a comparison of EvaluatorGPT for a more thorough evaluation.

2.2.1 Response time of DisasTeller

The initial response phase of real post-disaster scenarios typically occurs within the first 12 to 24 hours, focusing on immediate response actions such as deploying survey and assessment expert teams to generate the initial alert reports and public announcement [4,22]. The subsequent actions by emergency service teams to provide necessary assistance generally start within the first 72 hours, followed by the structured long-term recovery plan in 1 to 2 weeks [3,4]. Post-disaster response and recovery in real scenarios is a challenging and slow process due to the complexity of multi-stakeholder coordination and comprehensive damage assessment [6]. However, unlike human teams, taking several hours to communicate and intervene, from image inspection to disaster-relevant report generation, DisasTeller would only utilise around 4 minutes to coordinate all the agents and complete the given damage assessment tasks, thus simplifying the disaster response process and exhibit the potential to accelerate this process through the autonomous procedure.

2.2.2 Intermediate process tasks evaluation

The assessment of intermediate process tasks is crucial for each running of DisasTeller since the incorrect output of an individual agent would spread wrong information within the whole workflow and result in misleading final reports for all the stakeholders in the disaster event. This section evaluates two main intermediate tasks, including local disaster grading and alert map generation. Fig. 3 summarises the evaluation results across five rounds for both local disaster grading and alert map annotation tasks. The average score achieved by the framework for local disaster grading is 7.3/10, while for map annotations, the average score is 6.0/10. Each bar represents the score for one of the five rounds, demonstrating fluctuations in performance during the intermediate process tasks. This discrepancy between local disaster grading (7.3/10) and map annotation (6.0/10) exhibits the agent's weakness in integrating disaster-image-based and map-based data. Improving the LVLM agent's ability to process information from different sources could help address this gap.

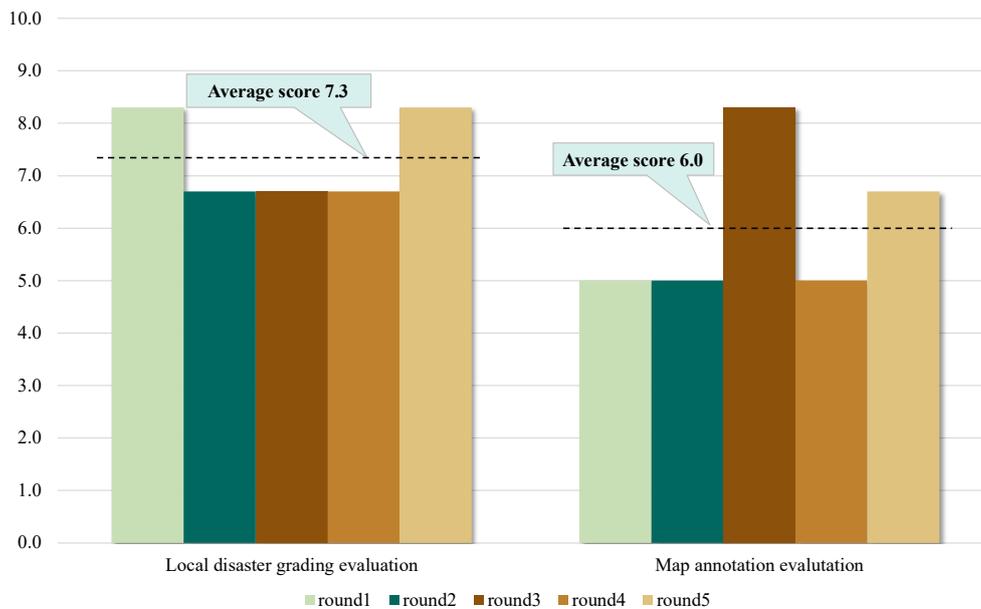

Fig. 3 The evaluation of intermediate process tasks in DisasTeller

2.2.3 Output reports evaluation

The output report quality of DisasTeller is crucial for post-disaster management because it would be a foundation for rapid and accurate decision-making. As presented in Fig. 4, the HumanEvaluator scores show more variability than EvaluatorGPT, reflecting the varying subject assessments for different persons. Reports like the future reconstruction plan, human allocation report and expert team report, which involve more technical, structured planning (budgeting, recovery, and restoration tasks), were rated higher by EvaluatorGPT, likely because the GPT model favours well-defined, structured tasks. On the other hand, reports that require more subjective judgment or social considerations, such as the public notice and alert team report, may have been better appreciated by human evaluators. Although DisasTeller's output reports achieve the matching score from the LVLM and human evaluators, further performance improvement is necessary for its integration into real-scenario disaster management workflow.

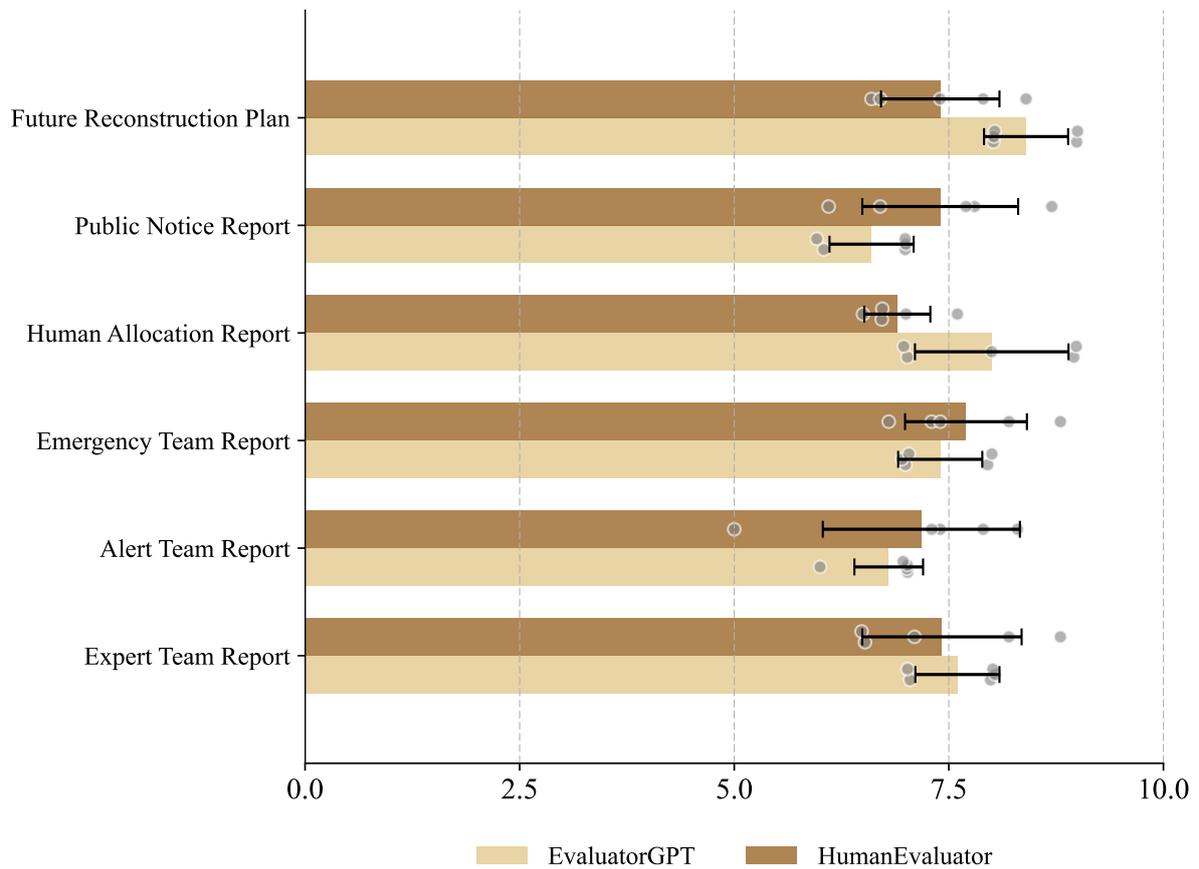

Fig. 4 The evaluation of output reports: the error bars represent the standard deviation, points in different colors mean the evaluation scores.

## 2.3 Limitations

Although DisasTeller exhibits impressive performance on autonomous post-disaster response tasks completion, integrating multi-LVLM agent-based engines like DisasTeller for post-disaster response still exists several inherent limitations and risks that need to be carefully mitigated to ensure effective and safe deployment.

One prominent concern is the risk of spreading inaccurate or misleading information (LVLM hallucination) during critical response phases. Despite the use of strict procedures, the reliance on LVLM-generated recommendations, if unchecked, can introduce flawed decision-making [20,23], potentially exacerbating the situation on the ground. Although some self-checking methods [24] are developed to mitigate the hallucination issues, concerns may still emerge when the LVLMs' limited ability to process novel disaster scenarios that fall outside their pre-trained data [23], which could lead to ineffective or inappropriate response strategies. To address this issue, integrating real-time data from trusted sources such as geographic information systems, satellite feeds, and sensor data can help multi-LVLM-powered engines like DisasTeller generate more accurate, contextually relevant responses for post-disaster scenarios and have up-to-date situational awareness [25]. In addition, setting up a peer-review system, such as reinforcement learning from human feedback [26], allows humans to cross-verify and provide corrective feedback on LVLM suggestions in real-time, could further enhance the reasoning reliability of LVLMs.

Since the sensitive information may be involved, safeguarding data privacy and ensuring robust security measures are vital when using LVLMs in post-disaster environments. Personal data (such as health records and identification details) and situational data (like locations and resource allocations) can be vulnerable to misuse if not properly protected. This could lead to privacy violations, identity theft, or even mistreatment of affected populations. Additionally, maintaining trust between responders and communities is critical, and any mishandling of data can undermine confidence in response efforts. Legal and ethical compliance is essential, requiring collaboration with legal experts and disaster authorities to align with data protection laws. Without proper security measures, sensitive data could be exposed to cybercriminals or used for misinformation [27], which could harm the disaster response process. Clear policies on data handling are essential to ensure responsible and secure use of LVLM-powered engines like DisasTeller in such environments.

In conclusion, while multi-LVLM-powered DisasTeller offers significant potential in disaster response, its limitations, such as LVLM hallucination, difficulty of handling novel disaster events, and data security concerns, must be diligently addressed to ensure their reliability and efficacy in critical situations. Yet, LVLMs are developing rapidly. For example, from GPT-3.5 to GPT-4, it takes only five months to improve the intelligence from a middle school or early high school student to a high school student or advanced undergraduate student [13], and the newly released GPT-o1 even exhibited a PhD level performance on specific tasks [28]. Fast ongoing technology upgrades and collaboration between developers, responders, and stakeholders will likely overcome these challenges and advance the safe use of LVLM-powered systems in disaster management.

## 3. Conclusion

This study develops the first multi-LVLM-agent-based framework DisasTeller for autonomous disaster response, aiming to accelerate the information sharing and team response in the disaster event, reduce disaster-relevant losses, and alleviate regional inequality in disaster response by its availability across developed and underdeveloped areas. Integrating the advanced reasoning ability of LVLMs with expert knowledge from external tools, DisasTeller can coordinate multiple LVLM agents and autonomously simulate various activities of human teams in the disaster event, including local on-site damage evaluation, alert map and disaster reports generation.

Assessment results showed that DisasTeller can significantly reduce the time needed for coordination and task completion compared to the current human-based disaster response procedure. Yet, its intermediate tasks, particularly local disaster grading and alert map generation, reveal performance discrepancies, indicating a weakness in integrating disaster-image and map-based data. The evaluations of its output reports have the matching results between LVLM and humans, demonstrating its potential for autonomously processing and sharing information among stakeholders in the disaster event.

However, the current outputs of DisasTeller are still limited by the ability of LVLMs and external tools, there is considerable room for improvement, especially in task-specific LVLM development and extra tools not restricted to the current operating framework. In addition, incorporating additional analysis of multimodal data from sensors and geography and real-time collaboration with human operators could further enhance DisasTeller's performance in complicate and real-world disaster scenarios.

In the evaluation process of DisasTeller outputs, the challenges come due to no specific indicators, but refining the experimental design could improve the reliability of evaluation results. Another critical challenge is the difficulty in reproducing the output of each independent run due to the random sampling [18] of LVLMs. Yet, although the randomness of the LVLM output can be controlled by the temperature parameter setting [19], the deterministic output would cause bias and reduce LVLM's problem-solving capability, particularly in disaster scenarios requiring diverse reasoning approaches [18]. Balancing LVLMs' output certainty and reasoning ability would be an open question that needs to be explored. Despite these challenges, our study highlights the promising potential of multi-LVLM-agent systems like DisasTeller to enhance disaster response significantly, improving efficiency and effectiveness across diverse disaster scenarios.